\documentclass[%
 reprint,
superscriptaddress,
nofootinbib,
 amsmath,amssymb,
 aps, preprintnumbers,
floatfix
]{revtex4-2}

\usepackage{graphicx}
\usepackage{dcolumn}
\usepackage{bm}
\usepackage{hyperref}
\usepackage[mathlines]{lineno}
\usepackage[caption=false]{subfig}
\usepackage{diagbox}
\usepackage{tabularx}
\usepackage{xcolor}

\usepackage{floatrow}
\usepackage{comment}

\usepackage{graphicx}
\usepackage[rightcaption]{sidecap}
\sidecaptionvpos{figure}{c}
 
\newfloatcommand{capbtabbox}{table}[][\FBwidth]

\usepackage{enumitem}

\usepackage{xspace}

\newcommand{\unit}[1]{\ensuremath{\mathrm{\,#1}}\xspace}
\newcommand{\units}[1]{\unit{#1}}
\newcommand{\e}{\unit{e^{-}}}

\begin{document}

\preprint{YITP-SB-2023-30, FERMILAB-PUB-23-0824-CSAID-PPD}

\title{SENSEI: First Direct-Detection Results on sub-GeV Dark Matter \texorpdfstring{\\}{} from SENSEI at SNOLAB
}

\author{Prakruth Adari}
\affiliation{\normalsize\it 
C.N.~Yang Institute for Theoretical Physics, Stony Brook University, Stony Brook, NY 11794, USA}
\affiliation{\normalsize\it 
Department of Physics and Astronomy, Stony Brook University, Stony Brook, NY 11794, USA} 

\author{Itay M. Bloch}
\affiliation{Berkeley Center for Theoretical Physics, University of California, Berkeley, CA 94720, U.S.A.}
\affiliation{Theoretical Physics Group, Lawrence Berkeley National Laboratory, Berkeley, CA 94720, U.S.A.}

\author{Ana M. Botti}
\affiliation{\normalsize\it 
Fermi National Accelerator Laboratory, PO Box 500, Batavia IL, 60510, USA}

\author{Mariano Cababie}
\affiliation{\normalsize\it 
Department of Physics, FCEN, University of Buenos Aires and IFIBA, CONICET, Buenos Aires, Argentina}
\affiliation{\normalsize\it 
Fermi National Accelerator Laboratory, PO Box 500, Batavia IL, 60510, USA}

\author{Gustavo Cancelo}
\affiliation{\normalsize\it 
Fermi National Accelerator Laboratory, PO Box 500, Batavia IL, 60510, USA}

\author{Brenda A. Cervantes-Vergara}
\affiliation{\normalsize\it 
Universidad Nacional Aut\'onoma de M\'exico, Ciudad de M\'exico, M\'exico}

\author{Michael Crisler}
\affiliation{\normalsize\it 
Fermi National Accelerator Laboratory, PO Box 500, Batavia IL, 60510, USA}

\author{Miguel Daal}
\affiliation{\normalsize\it 
 School of Physics and Astronomy, 
 Tel-Aviv University, Tel-Aviv 69978, Israel}

\author{Ansh Desai}
\affiliation{\normalsize\it 
Department of Physics and Institute for Fundamental Science, University of Oregon, Eugene, Oregon 97403, USA}

\author{Alex Drlica-Wagner}
\affiliation{\normalsize\it 
Fermi National Accelerator Laboratory, PO Box 500, Batavia IL, 60510, USA}
\affiliation{\normalsize\it Kavli Institute for Cosmological Physics, University of Chicago, Chicago, IL 60637, USA}
\affiliation{\normalsize\it  Department of Astronomy and Astrophysics, University of Chicago, Chicago IL 60637, USA}

 \author{Rouven Essig}
\affiliation{\normalsize\it 
C.N.~Yang Institute for Theoretical Physics, Stony Brook University, Stony Brook, NY 11794, USA}

 \author{Juan Estrada}
\affiliation{\normalsize\it 
Fermi National Accelerator Laboratory, PO Box 500, Batavia IL, 60510, USA}

\author{Erez Etzion}
\affiliation{\normalsize\it 
 School of Physics and Astronomy, 
 Tel-Aviv University, Tel-Aviv 69978, Israel}

\author{Guillermo Fernandez Moroni}
\affiliation{\normalsize\it 
Fermi National Accelerator Laboratory, PO Box 500, Batavia IL, 60510, USA}

\author{Stephen E. Holland}
\affiliation{\normalsize\it 
Lawrence Berkeley National Laboratory, One Cyclotron Road, Berkeley, California 94720, USA}

\author{Jonathan Kehat}
\affiliation{\normalsize\it 
 School of Physics and Astronomy, 
 Tel-Aviv University, Tel-Aviv 69978, Israel}

\author{Yaron Korn}
\affiliation{\normalsize\it 
 School of Physics and Astronomy, 
 Tel-Aviv University, Tel-Aviv 69978, Israel}
 
\author{Ian Lawson}
\affiliation{\normalsize\it SNOLAB, Lively, ON P3Y 1N2, Canada}

\author{Steffon Luoma}
\affiliation{\normalsize\it SNOLAB, Lively, ON P3Y 1N2, Canada}

 \author{Aviv Orly}
\affiliation{\normalsize\it 
 School of Physics and Astronomy, 
 Tel-Aviv University, Tel-Aviv 69978, Israel}

\author{Santiago E. Perez}
\affiliation{\normalsize\it 
Fermi National Accelerator Laboratory, PO Box 500, Batavia IL, 60510, USA}
\affiliation{\normalsize\it 
Universidad de Buenos Aires, Facultad de Ciencias Exactas y Naturales, Departamento de Física, Buenos Aires, Argentina}
\affiliation{\normalsize\it 
CONICET - Universidad de Buenos Aires, Instituto de Física de Buenos Aires (IFIBA). Buenos Aires, Argentina}

\author{Dario Rodrigues}
\affiliation{\normalsize\it 
Universidad de Buenos Aires, Facultad de Ciencias Exactas y Naturales, Departamento de Física, Buenos Aires, Argentina}
\affiliation{\normalsize\it 
CONICET - Universidad de Buenos Aires, Instituto de Física de Buenos Aires (IFIBA). Buenos Aires, Argentina}

\author{Nathan A. Saffold}
\affiliation{\normalsize\it 
Fermi National Accelerator Laboratory, PO Box 500, Batavia IL, 60510, USA}

\author{Silvia Scorza}
\affiliation{\normalsize\it 
Univ. Grenoble Alpes, CNRS, Grenoble INP, LPSC-IN2P3, Grenoble, 38000, France}

\author{Aman Singal}
\affiliation{\normalsize\it 
C.N.~Yang Institute for Theoretical Physics, Stony Brook University, Stony Brook, NY 11794, USA}
\affiliation{\normalsize\it 
Department of Physics and Astronomy, Stony Brook University, Stony Brook, NY 11794, USA}

 \author{Miguel Sofo-Haro}
\affiliation{\normalsize\it 
Fermi National Accelerator Laboratory, PO Box 500, Batavia IL, 60510, USA}
\affiliation{Universidad Nacional de C\'ordoba, CNEA/CONICET, C\'ordoba, Argentina}

\author{Leandro Stefanazzi}
\affiliation{\normalsize\it 
Fermi National Accelerator Laboratory, PO Box 500, Batavia IL, 60510, USA}

 \author{Kelly Stifter}
\affiliation{\normalsize\it 
Fermi National Accelerator Laboratory, PO Box 500, Batavia IL, 60510, USA}

\author{Javier Tiffenberg}
\affiliation{\normalsize\it 
Fermi National Accelerator Laboratory, PO Box 500, Batavia IL, 60510, USA}

\author{Sho Uemura}
\affiliation{\normalsize\it 
Fermi National Accelerator Laboratory, PO Box 500, Batavia IL, 60510, USA}

\author{Edgar Marrufo Villalpando}
\affiliation{\normalsize\it Kavli Institute for Cosmological Physics, University of Chicago, Chicago, IL 60637, USA}

\author{Tomer Volansky}
\affiliation{\normalsize\it 
 School of Physics and Astronomy,   
 Tel-Aviv University, Tel-Aviv 69978, Israel}

\author{Yikai Wu}
\affiliation{\normalsize\it 
C.N.~Yang Institute for Theoretical Physics, Stony Brook University, Stony Brook, NY 11794, USA}
\affiliation{\normalsize\it 
Department of Physics and Astronomy, Stony Brook University, Stony Brook, NY 11794, USA} 

\author{Tien-Tien Yu}
\affiliation{\normalsize\it 
Department of Physics and Institute for Fundamental Science, University of Oregon, Eugene, Oregon 97403, USA}

\collaboration{The SENSEI Collaboration }

\author{\vskip -4mm and Timon Emken}
\affiliation{\normalsize\it 
Oskar Klein Centre, Department of Physics, Stockholm University, Stockholm SE-10691, Sweden}

\author{Hailin Xu}
\affiliation{\normalsize\it 
C.N.~Yang Institute for Theoretical Physics, Stony Brook University, Stony Brook, NY 11794, USA}
\affiliation{\normalsize\it 
Department of Physics and Astronomy, Stony Brook University, Stony Brook, NY 11794, USA}

\date{\today}

\begin{abstract} 
We present the first results from a dark matter search using six Skipper-CCDs in the SENSEI detector operating at SNOLAB. We employ a bias-mitigation technique of hiding approximately 46\% of our total data and aggressively mask images to remove backgrounds. Given a total exposure after masking of 100.72 gram-days from well-performing sensors, we observe 55 two-electron events, 4 three-electron events, and no events containing 4--10 electrons.  The two-electron events are consistent with pileup from one-electron events.  Among the 4 three-electron events, 2 appear in pixels that are likely impacted by detector defects, although not strongly enough to trigger our ``hot-pixel'' mask. We use these data to set world-leading constraints on sub-GeV dark matter interacting with electrons and nuclei. 
\end{abstract}

\maketitle

\section{\label{sec:intro}Introduction}

Abundant astrophysical evidence indicates that a substantial fraction of the matter in our universe is composed of non-baryonic dark matter (DM). 
While many candidate particles with the required properties have been proposed, DM has yet to be detected in the laboratory. 
DM candidates with a mass of less than a proton (``sub-GeV DM'') are well-motivated and have increasingly been the target of experimental efforts in the last decade~\cite{Essig:2022dfa,Battaglieri:2017aum}. To date, the strongest direct-detection probes of DM with $\sim$eV-to-$\mathcal{O}(100)$~MeV masses are searches for electron recoils from DM-electron scattering~\cite{Essig:2011nj}, DM-nucleus scattering through the Migdal effect~\cite{Ibe:2017yqa} and bremsstrahlung~\cite{Kouvaris:2016afs}, and DM absorption on electrons~\cite{DIMOPOULOS1986145,PhysRevD.35.2752,An:2014twa,Bloch:2016sjj,Hochberg:2016sqx}. 

The Skipper Charge-Coupled Device (Skipper-CCD) technology can measure single-electron signals in a pixelated silicon target with high precision~\cite{Tiffenberg:2017aac}.
A particle interaction in silicon can excite an electron from the valence band to the conduction band in one of the pixels of the Skipper-CCD, which, given enough energy, can in turn create additional electron-hole pairs (below, simply referred to as ``electrons'' and denoted as ``$e^{-}$''). The charge in each pixel is then moved pixel-to-pixel to readout stages located in the corners of the Skipper-CCD, where the pixel charge is measured repeatedly and nondestructively to sub-electron precision. This technology paved the way for the development of SENSEI (\textit{Sub-Electron Noise Skipper-CCD Experimental Instrument}). The single-electron sensitivity corresponds to energy depositions of $\cal{O}$(1 eV), which are kinematically well-matched to the signals generated by the sub-GeV DM models described above. 
As a result, SENSEI~\cite{Crisler:2018gci,SENSEI:2019ibb,sensei2020} has begun to carve away at the allowed DM parameter space, as has DAMIC-M based on the same technology~\cite{DAMIC-M:2023gxo}. 

In this letter, we present the first results from a search for DM-electron scattering, DM-nuclear scattering through the Migdal effect, and dark photon DM absorption with the SENSEI apparatus operating in the SNOLAB underground laboratory, based on data taken from September 9, 2022 to April 10, 2023.   
Supplemental Materials (SM) contain additional details~\cite{SM-materials}.

\section{\label{sec:snolab}The SENSEI experiment at SNOLAB} 

\begin{figure}[t]
    \centering
    \includegraphics[width=1.0\textwidth]{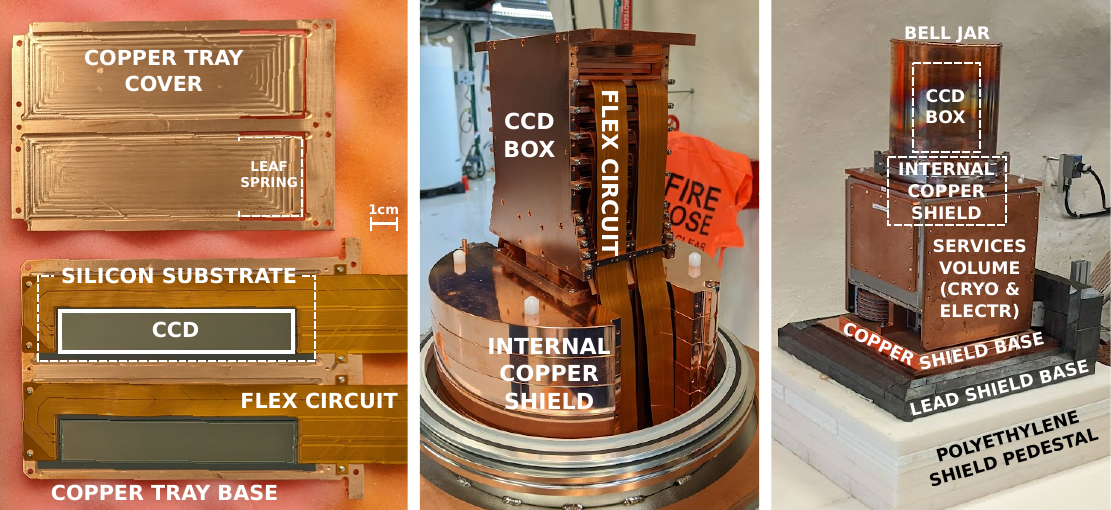}
     \caption{The SENSEI detector at SNOLAB. {\bf Left:} Two CCD modules in their copper tray. {\bf Middle:} Copper CCD box and trays deployed inside the vessel. {\bf Right:} Closed SENSEI vacuum vessel, before installing the outer copper, lead, and poly-water shields.}
    \label{fig:SENSEIdetector}
\end{figure}

Data were collected using six science-grade Skipper-CCDs designed at LBNL and fabricated at Teledyne DALSA Semiconductor.
These Skipper-CCDs, which were used in~\cite{skipper_at_surface,skipper_compton}, were fabricated on the same silicon wafers as those used in the previous SENSEI result~\cite{sensei2020} and have an identical design, except that their overall dimensions and packaging are slightly different.
The standard thickness is $675\units{\mu m}$, which, after repolishing, gives a thickness of $665\units{\mu m}$~\cite{4436592}. The active area of one CCD is $6144 \times 1024$ pixels of $15 \times 15\units{\mu m^2}$, for an active mass of 2.19~g.
In normal operation, each CCD is read out through four amplifiers, each of which reads a ``quadrant'' of 512 rows by 3072 columns. 
The CCD and a copper-Kapton flex circuit are glued to a silicon substrate and wirebonded. 
Two such modules are placed in a copper tray, which secures and cools the modules using a copper leaf spring (left panel in Fig.~\ref{fig:SENSEIdetector}).

The SENSEI apparatus at SNOLAB was installed in 2021. 
The experiment is contained in a vacuum vessel, which is separated into two sections.
The upper section holds a ``CCD box'' with capacity for up to 12~CCD trays (middle panel in Fig.~\ref{fig:SENSEIdetector}).
The CCD box is cooled through a cold finger, the temperature of which is maintained in a range of 125--145~K by a cryocooler.
The cooling power of the cryocooler declined with time, and was periodically recovered by warming the system to room temperature.
All major components of the upper section, including a 6-inch shield between the CCD box and the lower section, are made of oxygen-free high conductivity (OFHC) copper.
The lower section of the vacuum vessel holds the cryocooler and readout preamplifiers.  
The vacuum vessel (right panel in Fig.~\ref{fig:SENSEIdetector}) is shielded by a 2-inch inner layer of OFHC copper, a 3-inch middle layer of low-background lead~\cite{LBL}, and a 42-inch outer layer of water tanks and polyethylene (originally built for \cite{pico2L}).
The resulting rate of background events in the energy range from 500~eV to 10~keV is 140~events/kg/day/keV (see SM).

Each CCD is biased and read out by the Low Threshold Acquisition (LTA) readout electronics~\cite{Cancelo:2020egx}.
In order to reduce the ``spurious charge'' previously seen in these CCDs~\cite{sensei2020,sensei1e}, RC filters are used to shape the voltage waveforms applied to the CCD~\cite{janesick2001scientific}.
We collect 300 samples per pixel, with a single-sample readout time of 48.8\units{\mu s}.

We refer to one full readout of all six Skipper-CCDs as an ``image,'' and collect each image following the procedure outlined in~\cite{sensei1e}.
To prepare the CCD we adjust the bias voltages to erase the CCD and suppress dark current~\cite{holland_fullydepleted}, then repeatedly and rapidly (in 0.4~seconds/image) shift charge out of the CCD for three hours to remove charge released by long-lived traps~\cite{oscura_ccds}.
Then we expose the CCD for 20~hours, and read out every quadrant in parallel in 7.69~hours, which includes a prescan of 8 columns, an overscan of 120 columns, and a vertical overscan of 8 rows. 
Images are alternately designated as ``commissioning'' or ``hidden'' (either one commissioning followed by one hidden, or two commissioning followed by two hidden), restarting with a commissioning image after interruptions.
A total of 129 images (70 commissioning and 59 hidden) were collected. 
Only commissioning images are used to develop the analysis and data quality cuts.

\section{\label{sec:reco} Data reconstruction and selection} 

The averaging of Skipper samples, baseline subtraction, and clustering are performed as in~\cite{sensei2020}.
We define a pixel to have a \textit{thresholded} charge of (1\e, 2\e, $X$\e) if it has a \textit{measured} charge in the range ((0.7, 1.63], (1.63, 2.5], ($X-0.5$, $X+0.5$])\e, respectively. Given a pixel, a ``neighboring'' pixel is one of the eight adjacent pixels. A ``cluster'' or ``event'' is defined as a contiguous group of one or more neighboring pixels, each possessing a nonzero thresholded charge. The cluster charge is the sum of the thresholded pixel charges. 

Of the 24 total CCD quadrants, 12 were rejected based on their performance in the commissioning data. Three have low readout gain or high readout noise that prevents single-electron resolution.
Seven have high densities of single-electron events ($>5\times 10^{-4}$\e/pix) after applying all analysis cuts discussed in Section~\ref{sec:analysis} and averaging over all images, indicating poor CCD performance.
Two have multiple events in the 4--10 electron range after cuts, which can be linked to charge transfer inefficiencies that impact these quadrants. The remaining 12 ``good'' quadrants have readout noises in the range 0.131--0.153\e and single-electron event densities in the range (1.44--3.43)$\times 10^{-4}$\e/pix. 

We observe that images collected during periods of higher cold-finger temperature have correspondingly higher densities of single-electron events. To remove these images, we apply a ``hot image'' selection in two stages, looking only at regions of good quadrants that remain after applying the analysis cuts described in Section~\ref{sec:analysis}.
First, we treat each quadrant independently, and iteratively determine the set of images in which it is hot. 
A quadrant is marked hot in a given image if, considering the number of remaining pixels in this quadrant and image, as well as the density of single-electron events in this quadrant across all non-hot images, the Poisson $p$-value
meets a threshold such that if charge was uniformly distributed in time, the expected number of images with a good quadrant marked “hot” would be 0.5. 
Second, an image is considered ``good'' if no more than one of its quadrants is marked hot. 82 good images (45 commissioning and 37 hidden) survive this selection. 

The total exposure before cuts, in good quadrants and good images, is 534.89 g-days (293.53 commissioning, 241.35 hidden).

\section{\label{sec:analysis}Data analysis}

After determining good quadrants and good images, we select a clean sample of events by removing clusters from known background sources or with characteristics inconsistent with DM events by applying a series of analysis cuts (`masks').
Each mask flags specific pixels in each image for removal; unless otherwise specified, a masked pixel or cluster is still considered when computing each mask.
Each mask was developed to target a particular background source exclusively using the commissioning data, and was subsequently applied uniformly to both the commissioning and hidden data. 
Additionally, many of the masks are designed with specific statistical thresholds to remove only the most anomalous data, reducing the impact of bias.

We apply crosstalk, edge, serial register hit, bleeding zone, and halo mask algorithms that were previously 
used in~\cite{sensei2020}, but their parameters have been re-tuned for this analysis (see SM). Here we discuss the masks that are new or substantially different from~\cite{sensei2020}: 

\begin{itemize}[leftmargin=*]\addtolength{\itemsep}{-0.7\baselineskip}
\vspace{-3mm}
\item \textbf{Readout Noise.}
We observe time periods with a high level of readout noise from a source external to the CCD.
We mask rows where $>$6\% of pixels have measured charge more than $2\sigma$ away from an integer value, where $\sigma$ is the typical readout noise for the quadrant (see SM).

\item \textbf{Bad Pixels and Columns.}
Defects causing charge leakage or severe charge transfer inefficiency can create excess charge in specific pixels or columns, while amplifier light~\cite{sensei1e} can create a charge excess in columns near the readout amplifier.
We stack all images to identify and mask blocks of pixels and columns with a statistically improbable excess of charge. See SM for further discussion of block sizes and statistical tests.

\item \textbf{Full-Well Mask.}
We mask \textit{all} pixels in the same row as pixels whose charge is near the maximum capacity of the pixels (``full-well events''), as this results in severe charge trapping. 

\item \textbf{Low-Energy Cluster.}
After all other analysis masks, we still observe regions with excess charge due to amplifier light or inadequate masking.
We therefore mask pixels within $R_L=30$ pixels of any cluster, but not the cluster itself. 
Note that this removes all pairs of clusters less than $R_L$ apart from the analysis.
Detailed studies using data and simulations show that this mask is consistent with removing events in regions containing higher event densities from understood but inadequately masked detector backgrounds, without suppressing the density of DM-like events (see SM). This mask also prevents double-counting of DM events in non-contiguous pixels.

\end{itemize}

The fraction of pixels that survive masking ranges from 8.6--30.8\% across the good quadrants, and is 19.0\% overall.
The total exposure of unmasked pixels (referred to as the ``pixel exposure'') is 100.72 g-days (55.13 commissioning, 45.59 hidden).
See SM for cut efficiency and resulting pixel exposure for each subsequent mask.

Clusters are removed from the dataset if they have any pixel that has been masked.
We bin remaining events by cluster charge and pixel count; 2\e events are additionally binned by the spatial orientation of the cluster.

The 1\e channel is not used to constrain DM. Our data collection procedure was optimized for clusters containing 2\e--10\e, where 10\e is chosen arbitrarily above the point where our sensitivity becomes non-competitive with other experiments for all models we investigate.
We used the same exposure duration for all images and also did not independently measure exposure-independent sources of 1\e events such as spurious charge (as we did in~\cite{sensei2020, sensei1e}), and therefore we cannot isolate the fraction of the 1\e event density that corresponds to an event rate. 
Naively dividing counts by exposure for the quadrant with the lowest event density yields an upper bound on the rate of $(1.459\pm 0.020)\times 10^{-4}$~\e/pix/day, or $(419\pm 6)$~\e/g-day.

Observed event counts are shown in Table~\ref{tab:summary}.
We give event counts for both the hidden data and the combined commissioning+hidden data, since the hidden data was not statistically different from the commissioning data: 
a two-sided E-test~\cite{poisson_etest} finds that the event rates in the commissioning and hidden data for each shape are consistent, with the smallest $p$-value being $p=0.24$ for 2-pixel diagonal events.
In the range 11--100~\e, we observe one 26-\e event in the hidden data, consistent with the expected Compton-scattering background given the event rate in the 101--800~\e range. 

\begin{table}[t]
    \begin{center}
    \begin{footnotesize}
    \begin{tabularx}{\textwidth}{|X||c|c|c|c|c|c|c|}
        \hline 
        & \multicolumn{3}{c|}{Combined data} & \multicolumn{3}{c|}{Hidden data} & Reco. efficiencies \\
        \hline
        Bin & Ev. & Bkgd. & Expo. & Ev. & Bkgd. &  Expo. & Shape/ID/Geom. \\
        \hline 
        2e2p, h & 10 & 10.66 & 13.58 & 6 & 4.75 & 6.12 & 0.15/0.96/0.94 \\
        2e2p, v & 13 & 12.65 & 16.21 & 6 & 5.61 & 7.36 & 0.18/0.96/0.94 \\
        2e2p, d & 32 & 25.31 & 16.82 & 18 & 11.22 & 7.65 & 0.19/0.96/0.93 \\
        \textbf{2e, all} & \textbf{55} & \textbf{48.62} & \textbf{46.61} & \textbf{30} & \textbf{21.57} & \textbf{21.13} & \textbf{0.51/0.96/0.94} \\
        3e2p & 3 & 0.01 & 30.87 & 2 & 0.01 & 14.01 & 0.33/0.98/0.96 \\
        3e3p & 1 & 0.06 & 26.84 & 0 & 0.03 & 12.21 & 0.31/0.94/0.90 \\
        \textbf{3e, all} & \textbf{4} & \textbf{0.07} & \textbf{57.71} & \textbf{2} & \textbf{0.03} & \textbf{26.22} & \textbf{0.64/0.96/0.93} \\
        4e2p & 0 & 0.00 & 19.51 & 0 & 0.00 & 8.85 & 0.21/0.98/0.96 \\
        4e3p & 0 & 0.00 & 27.60 & 0 & 0.00 & 12.55 & 0.31/0.96/0.92 \\
        4e4p & 0 & 0.00 & 15.93 & 0 & 0.00 & 7.25 & 0.19/0.93/0.88 \\
        \textbf{4e, all} & \textbf{0} & \textbf{0.00} & \textbf{63.03} & \textbf{0} & \textbf{0.00} & \textbf{28.66} & \textbf{0.71/0.96/0.92} \\
        5e, all & 0 & 0.00 & 65.56 & 0 & 0.00 & 29.84 & 0.75/0.95/0.91 \\
        6e, all & 0 & 0.00 & 67.31 & 0 & 0.00 & 30.63 & 0.78/0.95/0.90 \\
        7e, all & 0 & 0.00 & 68.53 & 0 & 0.00 & 31.19 & 0.80/0.95/0.89 \\
        8e, all & 0 & 0.00 & 69.52 & 0 & 0.00 & 31.67 & 0.82/0.95/0.89 \\
        9e, all & 0 & 0.00 & 70.30 & 0 & 0.00 & 32.02 & 0.84/0.94/0.88 \\
        10e, all & 0 & 0.00 & 70.89 & 0 & 0.00 & 32.30 & 0.85/0.94/0.88 \\
        \hline
    \end{tabularx}
    \vspace{-3mm} 
    \end{footnotesize}
    \caption{Observed events after masking, expected background, and effective exposure for the combined data and the hidden portion only, as well as reconstruction efficiencies. Events are  binned by charge (the number preceding ``e'') and pixel count (the number preceding ``p''); 2\e events are further binned by spatial orientation ({\it h}orizontal, {\it v}ertical or {\it d}iagonal). Totals for $2-4$\e events are given for the reader's convenience and not used in the extraction of the exclusion limit. Observed events (``Ev.'') and expected background (``Bkgd.'') are counts; effective exposures (``Expo.'') are in gram-days. The reconstruction efficiency is decomposed into cluster shape (``Shape''), misidentification (``ID''), and geometric (``Geom.'') factors. \label{tab:summary}
    \vspace{-3mm}}
    \end{center}
\end{table}

We compute the expected number of 2\e, 3\e, and 4\e background events due to pileup of 1\e events in each quadrant of each image. 
These values are given for each bin in Table~\ref{tab:summary}.
This calculation is done by taking the Poisson probability of an N-fold coincidence given the density of 1\e events and the number of allowed cluster shape permutations, and multiplying by the number of unmasked pixels.
This underestimates the pileup rate and gives a conservative limit if the 1\e event density is not uniform within an image (see SM).
Our calculation is subject to fluctuations in the number of 1\e events and the resulting relative uncertainty on the expected background is 0.9\% in the worst case (for diagonal 2\e events).

The effective exposure is calculated via Monte Carlo sampling for each bin, and can be thought of as the pixel exposure (which is calculated after masking) multiplied by the efficiency to reconstruct a cluster of the size and shape of the given bin. The simulation takes into account three effects that impact the reconstruction efficiency, which are described in subsequent paragraphs.

The first effect is cluster shape, which is heavily impacted by diffusion of charge through the CCD. Diffusion is simulated using the same model and parameters as in~\cite{sensei2020}, since the CCD properties and bias voltage are identical. Due to the charge dependence of the diffusion width~\cite{sensei_diffusion}, this model, fitted on muon tracks, is known to overestimate diffusion. 
Using high-energy clusters from SNOLAB data, we measure the maximum observed spread, $\sigma_{\rm max}$, as a function of cluster charge and confirm that our model overestimates $\sigma_{\rm max}$ by at most $5\%$ for low-energy clusters.  This results in a conservative underestimate for the efficiency of low-energy DM events to produce contiguous pixels by 2--3\%, which was confirmed with simulations. 

Using this diffusion model, we simulate events with a uniform depth distribution and determine which remain contiguous.
Of the contiguous events, we remove several specific shapes due to high background levels. We observe 9 (21) single-pixel 2\e events in commissioning (combined) data, which is a significant excess over the pileup expectation of 3.52 (6.33) events.
We do not observe an excess of two-pixel 2\e events.
The source of this excess is unknown, but we speculate that it may be associated with unmasked hot pixels, amplifier light (multiple photons emitted in bursts, or single photons above the 2\e threshold), or other stray photons above the 2\e threshold that are absorbed near the CCD frontside.\footnote{The 5 single-pixel 2\e events observed in~\cite{sensei2020} were also in excess of the pileup expectation.} 
To avoid such possible backgrounds, we remove \textit{all} single-pixel multi-\e events.\footnote{We note, however, that we do not observe any single-pixel events containing more than 2\e 
in the entire dataset.} 
We also observe an excess of 2\e and 3\e ``horizontal'' clusters in two of the good quadrants across several images (178 2\e-events versus 1.99 expected from pileup, and 6 3\e events versus $\sim$0 expected), where several pixels next to each other in the same row contain 1\e. These are likely associated with charge transfer inefficiencies when draining charge from the readout amplifier. We remove all horizontal clusters in these two quadrants, and all horizontal clusters with 3\e or above in any quadrant. 

We call the combination of diffusion and specific shape removal the ``cluster shape efficiency,'' i.e.~the probability to obtain a contiguous cluster that passes the additional cluster shape cuts, given an initial event charge.


\begin{SCfigure*}[0.37][t!]
    \centering
    \caption{
\textbf{Top left:} Solid (dashed) cyan and olive lines are the 90\%~C.L.~constraints on DM-\e cross section, $\overline{\sigma}_e$, 
versus DM mass, $m_\chi$, for light mediators from the hidden (hidden + commissioning) data. Cyan (olive) is for halo DM-\e scattering (solar-reflected DM, assuming a dark photon mediator). Freeze-in line (orange) is from~\cite{Essig:2011nj,Essig:2015cda,Chu:2011be,Dvorkin:2019zdi}. 
Other bounds are from SENSEI~\cite{SENSEI:2019ibb,sensei2020} and~\cite{Aprile:2019xxb,DAMIC-M:2023gxo,Emken:2024nox}. 
\textbf{Top right:} As for top-left but for DM-\e scattering through a heavy mediator; other bounds from~\cite{SENSEI:2019ibb,sensei2020,Aprile:2019xxb,DarkSide:2022knj,DAMIC-M:2023gxo,DAMIC-M:2023hgj,Emken:2024nox}. 
Benchmark targets from~\cite{Essig:2022dfa,Boehm:2003hm, Essig:2011nj,Lin:2011gj,Essig:2015cda,Hochberg:2014dra,Kuflik:2017iqs,DAgnolo:2019zkf} are in orange. 
\textbf{Bottom left:} 
 Bounds on the DM-nucleon cross section, $\overline{\sigma}_n$, for a heavy mediator, using the Migdal effect~\cite{sensei2020,Berghaus:2022pbu,Aprile:2019jmx,Essig:2019xkx,DarkSide:2022dhx}, except for solid black lines, which assume elastic scattering~\cite{SuperCDMS:2020aus,Abdelhameed:2019hmk}. 
\textbf{Bottom right:} 
 Bounds on the kinetic-mixing parameter, $\epsilon$, versus the dark-photon mass, $m_{A'}$, for dark-photon-DM absorption. Others bounds are from~\cite{Bloch:2016sjj,Crisler:2018gci,SENSEI:2019ibb,sensei2020,An:2013yfc,Redondo:2013lna,XENON:2021qze,PandaX:2023xgl}. 
 \label{fig:DMLimits}
 }
\begin{minipage}{0.7\textwidth}
\begin{center}
\includegraphics[width=0.49\textwidth]{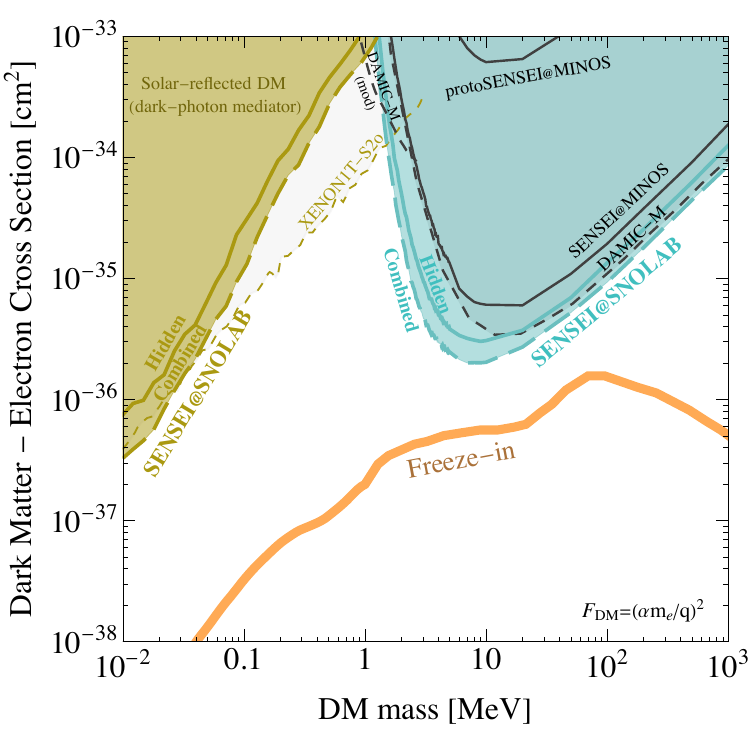}
 \includegraphics[width=0.49\textwidth]{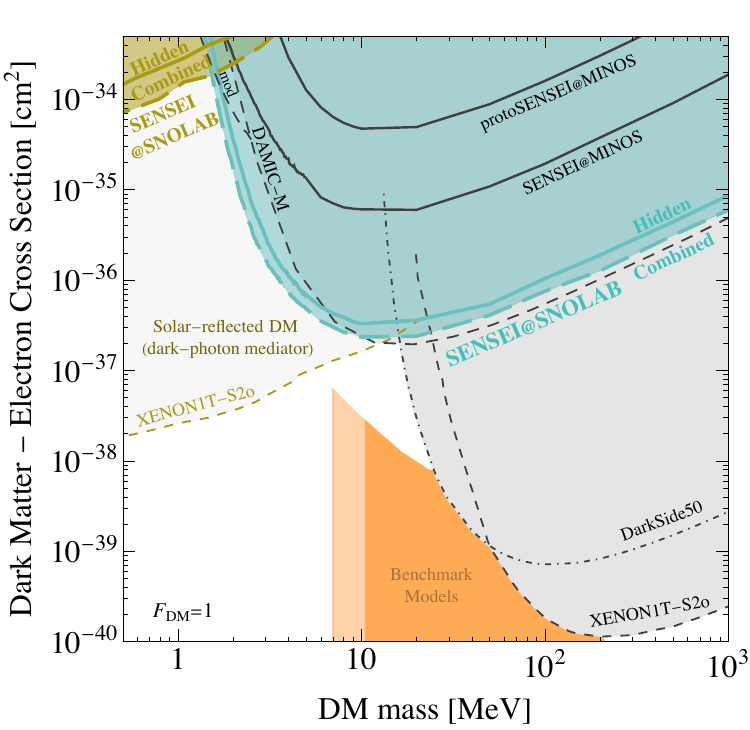}\\
\includegraphics[width=0.49\textwidth]{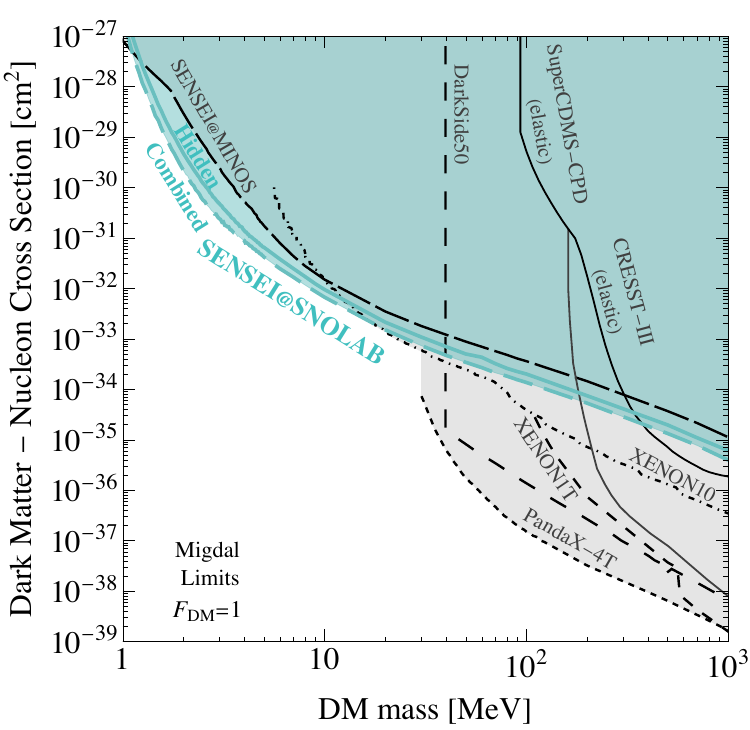}
\includegraphics[width=0.49\textwidth]{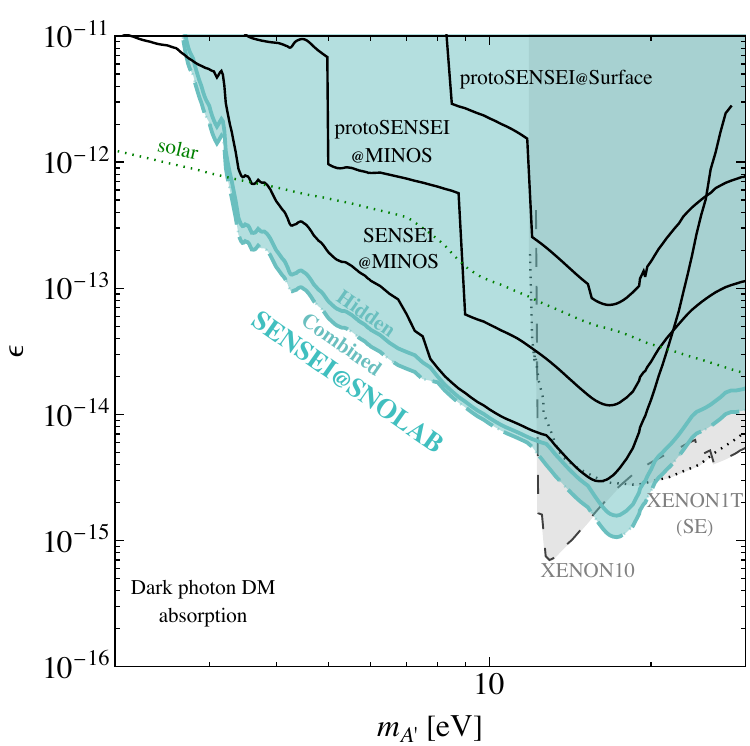}
\end{center}
\end{minipage}
\end{SCfigure*}

We use a Monte Carlo sampling technique to estimate the total exposure across the CCD, where each pixel has a unique exposure due to different readout times.  Simulated events are uniformly injected in all images, and the particular exposure contribution for each event (based on the pixels that it lands in) is weighted by the analytical probability for every pixel of the event to have its charge correctly reconstructed (the ``misID efficiency''). This efficiency calculation is a conservative underestimate since it assumes a misreconstructed event is removed or lost, when in fact it merely changes the cluster charge. We then sum the weighted exposure from each event that survives masking and normalize by the total number of pixels in the CCD and original number of events, resulting in the effective exposure.
In this way, we can define the ``geometric efficiency'', meaning the probability for an event whose center is unmasked to have no pixels overlapping with a masked region~\cite{senseiMariano}. 

Following the calculation of effective exposure, we can back out the impact of each effect that was simulated, resulting in our reconstruction efficiencies.
The effective exposure and reconstruction efficiencies for each bin are given in Table~\ref{tab:summary}. The statistical uncertainty of these simulations contribute a relative uncertainty to the effective exposure of 0.7\% in the worst case (for horizontal/vertical 2\e events).

\section{\label{sec:results}Dark matter results} 

As summarized in Table \ref{tab:summary}, we observe 30 (55) 2\e-events in the hidden (combined) data and no 4\e to 10\e-events, consistent with pileup background expectations.  We do observe, however, two (four) 3\e-events, expecting only 0.03 (0.07) pileup events, in the hidden (combined) data.
The 3\e events are unlikely to arise from Compton scattering~\cite{skipper_compton, DAMIC-M:2022xtp,Essig:2023wrl} or the partial collection of charge from higher-energy events~\cite{damic_background}, as these would also produce many events with higher charge.
They are also inconsistent with any external origin creating 3\e events at either CCD-surface: front-side events would usually form single-pixel clusters while we observe none among the rejected events, and a back-side source would need to first penetrate an opaque layer of in-situ doped polysilicon. 
However, two of these 3\e-events (both in the hidden data) occur in positions where we see an excess number of events in stacked images (with probabilities of $2.7\times 10^{-3}$ and $2.0\times 10^{-6}$), albeit not enough to have triggered the bad-pixel mask, suggesting that they are due to CCD defects creating localized charge leakage. Since this was determined only after opening the hidden data, we do not remove these from our analysis. 

Using the data in Table~\ref{tab:summary}, we show the  90\%~C.L.~limit constraints in Fig.~\ref{fig:DMLimits} (cyan lines/regions) on halo DM that scatters off electrons for heavy and light mediators~\cite{Essig:2011nj}, DM that scatters off nuclei through the Migdal effect for heavy mediators~\cite{Ibe:2017yqa}, 
and DM that is absorbed by electrons~\cite{DIMOPOULOS1986145,PhysRevD.35.2752,An:2014twa,Bloch:2016sjj,Hochberg:2016sqx}.  
Assuming a dark-photon mediator, we also show bounds on the ``solar-reflected DM'' component (olive)~\cite{An:2017ojc,Emken:2017hnp,Emken:2019hgy,Emken:2021lgc,An:2021qdl,Emken:2024nox}.  Solid (dashed) lines use the hidden (combined) data, with the dashed lines facilitating comparison with results that are not based on a hidden search. 
Our bounds improve on previous bounds~\cite{Crisler:2018gci,SENSEI:2019ibb,sensei2020,Angle:2011th,Essig:2012yx,Aprile:2016wwo,DAMIC:2016qck,Essig:2017kqs,Agnese:2018col,Agnes:2018oej,Blanco:2019lrf,Aprile:2019xxb,Aguilar-Arevalo:2019wdi,Essig:2019xkx,Aprile:2019jmx,DAMIC:2019dcn,Amaral:2020ryn,Arnaud:2020svb,XENON:2021qze,PandaX-II:2021nsg,SuperCDMS:2022kgp,EDELWEISS:2022ktt,DAMIC-M:2023gxo,SuperCDMS:2023sql} for a range of DM models, especially for DM interacting with electrons through a light mediator and for dark-photon DM absorption, but also for DM-nuclear interactions below $\sim$30~MeV for heavy mediators. In the SM, we show the limit on DM that scatters off nuclei through the Migdal effect for light mediators, which improves on previous bounds for DM masses below $\sim$40~MeV. 

As in~\cite{sensei2020}, we use a likelihood-ratio test based on~\cite{Cowan:2010js} and a toy MC to compute the distribution of the test statistics used for the calculation of the $p$-value, the one difference being that we account for the expected pileup background in each bin as a known background. See SM for the full statistical treatment.

The DM-electron scattering rates are from {\tt QEDark}~\cite{QEdark,Essig:2015cda}. A comparison to the DM-electron scattering calculations with {\tt DarkELF}~\cite{Knapen:2021bwg}, {\tt EXCEED-DM}~\cite{Griffin:2021znd,Trickle:2022fwt}, and {\tt QCDark}~\cite{Dreyer:2023ovn,QCDark} is given in the SM.  
We use the Migdal rates from~\cite{Berghaus:2022pbu}, which uses input from DarkELF~\cite{Knapen:2021bwg,Knapen:2020aky}, and the absorption rates from~\cite{HENKE1993181,EDWARDS1985547}. 
For solar-reflected DM, we use the simulations from~\cite{Emken:2024nox} (using~\cite{Emken:2019hgy,Emken:2021lgc,DaMaSCUSsun}) and the DM-electron scattering cross section from~\cite{Essig:2024ebk}. 
We use the DM halo parameters in~\cite{Baxter:2021pqo}. We follow the ionization model in~\cite{Ramanathan_2020} to compute the number of electrons produced by energy deposits in the CCD. 
The previous SENSEI limit~\cite{sensei2020} has been recast with these halo parameters and ionization model. 

We also compute model-independent rates of events for each charge $n$, $R_{n\! \e}$ (all following rates are in events/g-day). 
The maximum-likelihood $R_{2e^-}$ in the combined (hidden-only) data is $8.57\times 10^{-2}$ ($3.18\times 10^{-1}$), and the 90\% C.L. upper limit is $3.25\times 10^{-1}$ ($7.32\times 10^{-1}$), a factor of $\sim$14 lower than our previous limit~\cite{sensei2020}.
The maximum-likelihood $R_{3e^-}$ in all (hidden) data is $6.85\times 10^{-2}$ ($7.56\times 10^{-2}$), and the 90\% C.L. upper limit is $1.49\times 10^{-1}$ ($1.97\times 10^{-1}$).
For higher charges where we observe no events, the maximum-likelihood rate is zero and the upper limit is 2.30
events divided by effective exposure.

In summary, we presented world-leading constraints from the SENSEI detector at SNOLAB.  The observed 2\e events are consistent with pileup expectations from 1\e events.  Four 3\e events are observed, with two of them later found to occur in pixels with an excess number of events in stacked images. The SNOLAB setup has been upgraded with additional detectors for another science run. A larger dataset will allow us to further investigate if the observed 3\e events are consistent with excesses in stacked images. A dedicated run will measure the 1\e event rate.  Further improvements to the data-taking and analysis based on what was learned after opening the hidden data are expected to further increase our sensitivity to low-mass DM.

\begin{acknowledgments}
We thank Ryan Plestid for collaboration on the DM-electron scattering rates for solar-reflected DM to appear in~\cite{Essig:2024ebk}. We are grateful for the support of the Heising-Simons Foundation under Grant No.~79921.
This document was prepared by the SENSEI collaboration using the resources of the Fermi National Accelerator Laboratory (Fermilab), a U.S. Department of Energy, Office of Science, Office of High Energy Physics HEP User Facility. Fermilab is managed by Fermi Research Alliance, LLC (FRA), acting under Contract No.~DE-AC02-07CH11359.
We would like to thank SNOLAB and its staff for support through underground space, logistical and technical services. SNOLAB operations are supported by the Canadian Foundation for Innovation and the Province of Ontario, with underground access provided by Vale at the Creighton mine site.
The CCD development work was supported in part by the Director, Office of Science, of the DOE under No.~DE-AC02-05CH11231. RE acknowledges support from DOE Grant DE-SC0009854 and Simons Investigator in Physics Award~623940, which also provide support for AS and HX. 
TV is supported, in part, by the Israel Science Foundation (grant No. 1862/21), 
by the NSF-BSF (grant No.\ 2021780) and by the European Research Council (ERC) under the EU Horizon 2020 Programme (ERC-CoG-2015, Proposal No.\ 682676 LDMThExp).
RE and TV acknowledge support from the Binational Science Foundation (grant No.\ 2020220). 
IB is grateful for the support of the Alexander Zaks Scholarship, The Buchmann Scholarship, and the Azrieli Foundation.
The U.S. Government retains and the publisher, by accepting the article for publication, acknowledges that the U.S. Government retains a non-exclusive, paid-up, irrevocable, world-wide license to publish or reproduce the published form of this manuscript, or allow others to do so, for U.S. Government purposes.
\end{acknowledgments}

\newpage 

\begin{center}
    \large{\textbf{Supplemental Materials}}
\end{center}

\setcounter{section}{0}

In these Supplemental Materials, we provide additional details and information to clarify the main results.  We present the high-energy event spectrum in Section~\ref{app:HE-spectrum}; provide more details in Section~\ref{app:mask-details} on the masks that were previously used in~\cite{sensei2020} but were re-tuned for this analysis; provide more details on the low-energy cluster mask in Section~\ref{app:Low-E-Mask}; detail the efficiencies of applying each mask to the data in Section~\ref{app:cut-flow}; describe our limit-setting procedure in Section~\ref{app:likelihood}; compare the DM-electron scattering limits using different calculations for the cross section in Section~\ref{app:comparison}; compare the DM-electron scattering limits using the combined bins to the limits using individual bins in Section~\ref{app:combined-vs-individual}; and give the limit on DM-nucleus scattering via a light mediator from the Migdal effect in Section~\ref{app:Migdal-light}. 

\section{High-Energy Event Spectrum}\label{app:HE-spectrum}
\begin{figure}[tbh]
    \centering
    \includegraphics[width=\textwidth]{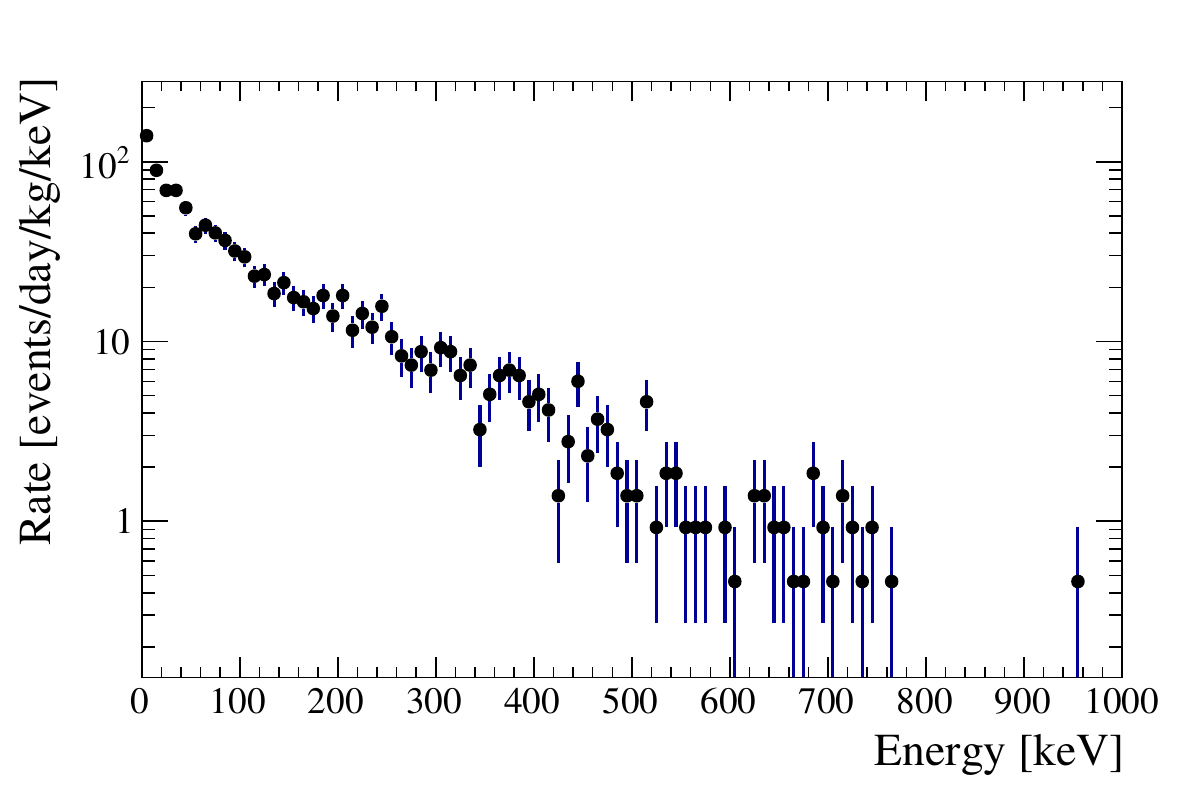}
    \caption{High-energy event spectrum from 500~eV to 1~MeV, using all good quadrants and images from the combined data. A subset of masks have been applied to remove instrumental backgrounds.}
    \label{fig:DRU}
\end{figure}

The high-energy event spectrum from 500~eV to 1~MeV is shown in Fig.~\ref{fig:DRU}.  In the energy range from 500~eV to 10~keV, the background event rate is $\sim$140~events/kg/day/keV, which is a factor of $\sim$24 smaller than the background rate of the SENSEI detector located in the MINOS cavern~\cite{sensei2020}. We did not perform a surface etch of our copper components, which is expected to reduce this background further.

\section{Additional Mask Details}\label{app:mask-details}

Several of the masks used in our analysis have been employed previously in~\cite{sensei2020}, but their parameters have been re-tuned for this analysis. We discuss these previously-used masks below, which target crosstalk, edge events, serial register hits, bleeding zones, and ``halos'' around high-energy clusters. We also give additional technical details on several of the masks described in the main text.

\begin{itemize}[leftmargin=*]\addtolength{\itemsep}{-0.7\baselineskip}
\item \textbf{Readout Noise.}
As described in the main text, this mask removes events due to readout noise coming from a source external to the CCD. The ``typical readout noise'' level is determined separately for each quadrant. First, we apply a cleanup cut to remove extremely noisy rows where more than 5\% of the pixels have $|\text{measured charge}|>200$ analog-to-digital units (approximately $1/3$\e). 
We then fit the cores of the 0\e and 1\e Gaussian peaks to obtain the typical readout noise in the quadrant, which is used in the Readout Noise mask as described in the main text. This process has a $6.9\times 10^{-5}$ probability of removing a ``good row'' due to fluctuations of the measured charge.

\item \textbf{Crosstalk.}
A pixel is masked if it is read at the same time as another pixel (on a different quadrant) containing $>$700 electrons, which can produce fake signals due to electronic crosstalk.

\item \textbf{Edge Mask.}
We remove 30 pixels around all edges of a quadrant, which corresponds to applying the Low-Energy Cluster mask for any possible clusters located just outside of the quadrant.

\item \textbf{Bad Pixels and Columns.}
As described in the main text, we stack all images to identify regions of the CCDs with a statistically improbable excess of
charge. Block sizes of 1x1-5x5 pixels and 2-5 columns are checked, which covers the size scales relevant to diffusion from dark spikes. Then, we also check blocks of columns equivalent to 1/16 of a quadrant, which is the typical size scale for amplifier light. The statistical significance of excess charge is determined by setting thresholds on Poisson $p$-values. For each block size, the $p$-value threshold is set such that if charge was uniformly distributed, the expected number of blocks removed would be 0.5. In the particular case of the column blocks of 1/16 of a quadrant, the $p$-value threshold is set such such that the expected number of columns removed is 0.5. Due to the fact that this mask targets columns or pixels that are consistently hot, if a pixel or column is removed in this mask, it is removed in every image.

\item \textbf{Serial Register Hit.} 
An event that hits the serial register can create an isolated horizontal line of charge~\cite{skipper_at_surface}.
The left and right tails of a wide, low-charge serial register hit can contain disconnected few-\e clusters, which constitute a background for DM searches.
This mask is designed to remove these serial register hits and their tails without removing normal, compact clusters. 
In each row, we examine all possible windows of 30 consecutive pixels.
If a window contains 3 or more nonempty pixels with $\geq$5 pixel separation between the first and last, and the 30-pixel windows immediately above and below the window in question each contain $\leq$1 nonempty pixels, we mask both the window in question as well as the 60 pixels to both its left and right. 

\item \textbf{Bleeding Zone Mask.}
To mask spurious events from charge transfer inefficiencies, we mask all (50) pixels upstream (later in the readout direction) in the vertical (horizontal) direction 
of any pixel containing more than 100\e.

\item \textbf{Halo Mask.} 
High-energy background events correlate with an increased rate of low-energy events in nearby pixels~\cite{sensei2020,Du_2022,Du:2023soy}. 
We mask pixels within 60~pixels of ``high-energy'' pixels, defined as those with more than 100\e.

\end{itemize}

\section{Discussion of Low-Energy Cluster Mask}\label{app:Low-E-Mask}

The low-energy cluster mask is designed to remove regions of images with excess events. In practice, we see that it suppresses two main types of effects.

\begin{figure}[tbh]
    \centering
    \includegraphics[width=\textwidth]{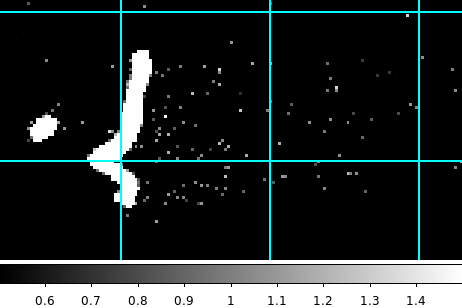}
    \caption{Charge transfer inefficiency creates a tail of charge to the right of this high-energy event. The 50-pixel spacing of the cyan grid lines equals the range of the bleeding zone mask, and shows that much of the tail is not masked.}
    \label{fig:bleed_edge}
\end{figure}

First, if a high-energy event creates a region with a very high local density of 1\e events that is inadequately masked by other masks, the low-energy cluster mask will remove this region.
Figure~\ref{fig:bleed_edge} shows one example, where horizontal charge transfer inefficiency creates a tail of charge that extends beyond the range of the bleeding zone mask that was applied to this cluster.
This effect and other masking failures were not identified in the tuning of masks on commissioning data due to the small sample size. Still, the possibility of inadequate masking of rare effects was anticipated, and was a main motivation for the low-energy cluster mask.

Second, the 1\e event density varies within a quadrant due to effects such as amplifier light.
These variations are not completely suppressed by the bad column mask, which has a strict statistical threshold intended to minimize the chance of masking a region due to random fluctuations in the event density.
The low-energy cluster mask further suppresses these variations, which improves the fidelity of our pileup background estimate.

\begin{figure}[tbh]
    \centering
    \includegraphics[width=0.83\textwidth]{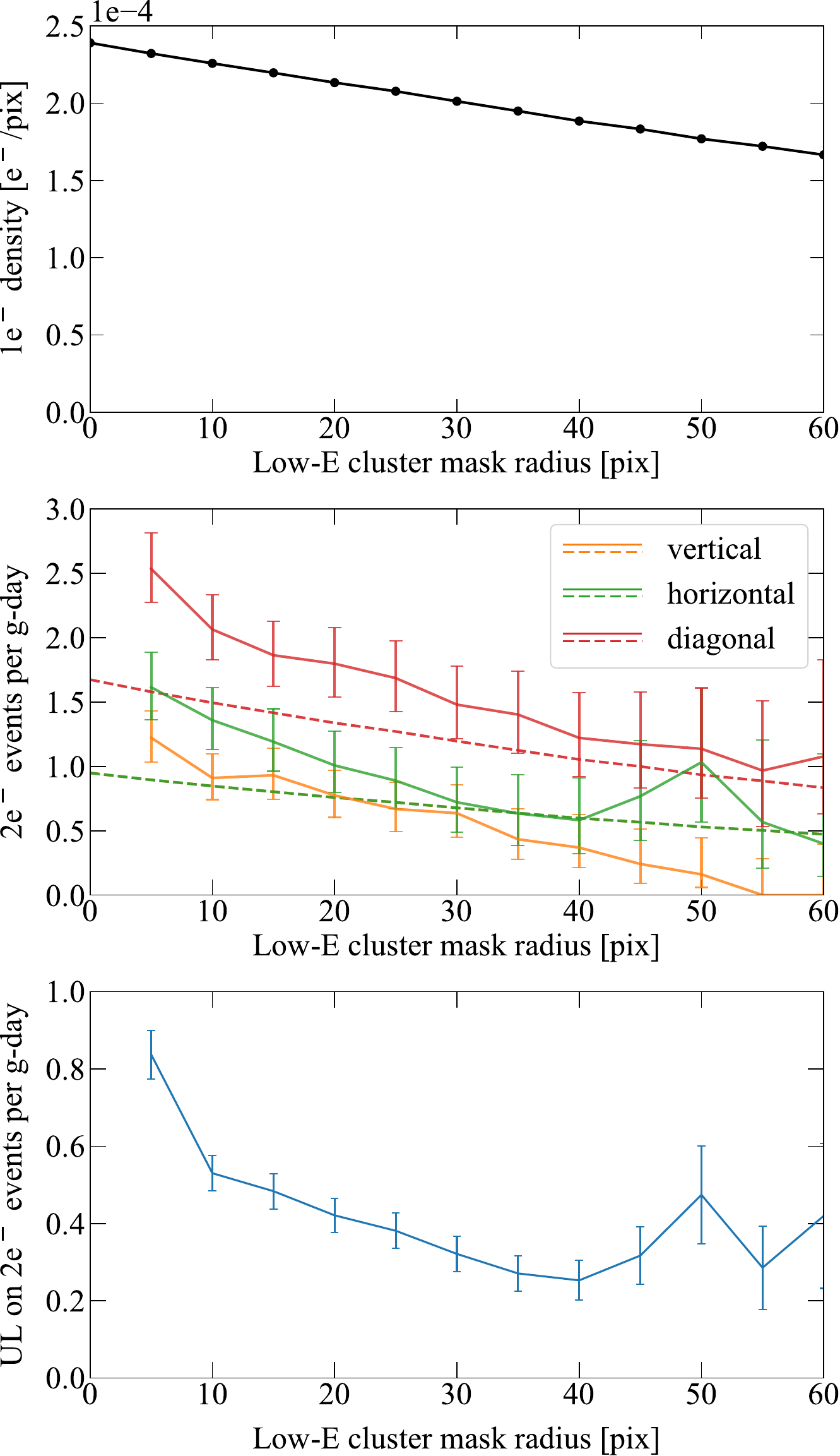}    
    \caption{\textbf{Top:} 1\e event density as a function of low-energy cluster mask radius. \textbf{Middle:} Rates of 2\e events as a function of low-energy cluster-mask radius. Solid lines indicate measured rates with the error bars corresponding to a 68\% confidence interval~\cite{Feldman_1998}, dashed lines indicate expected background rates of 2\e events due to pileup assuming uniform 1\e event density within each quadrant of each image. Note that the expected pileup rates for vertical and horizontal events are identical, so those dashed lines overlap. \textbf{Bottom:} Upper limit of 90\% confidence interval on combined 2\e events per g-day as a function of low-energy cluster mask radius. Error bars correspond to relative error on the number of observed counts, assuming Poisson statistics. Increasing error bars at large low-E cluster radii are due to small statistics as more events are masked out. All plots use the combined data (commissioning+hidden).}
    \label{fig:LowEClusterEvents}
\end{figure}

We use both data and Monte Carlo to study the effect of the low-energy cluster mask.
As shown in Fig.~\ref{fig:LowEClusterEvents}, the measured densities of 1\e and multi-\e events vary depending on the low-energy cluster mask radius $R_L$. This is due to variations in the 1\e density between and within quadrants, which has the following effects:
\begin{itemize}[leftmargin=0cm,itemindent=.5cm,labelwidth=\itemindent,labelsep=0cm,align=left]\addtolength{\itemsep}{-0.6\baselineskip}%
    \item Since the 2\e background from pileup is proportional to the square of the local 1\e density, the true number of 2\e pileup events is likely to be greater than our estimate, which conservatively assumes a uniform 1\e density across a given quadrant and image.
    \item The efficiency of the low-energy cluster mask depends on the local 1\e density, with regions with a higher 1\e density being more heavily masked.
    \item As $R_L$ increases, the quadrants and regions with a higher density of 1\e events become more heavily masked and the overall 1\e event density decreases (top panel of Fig.~\ref{fig:LowEClusterEvents}). The measured 2\e rate  decreases and approaches the pile-up estimate (middle panel of Fig.~\ref{fig:LowEClusterEvents}).
\end{itemize}

An increasing value of $R_L$ gives improved agreement between observed and expected 2\e counts due to removal of backgrounds, at the expense of exposure loss. At $R_L=30$, which is used in this analysis, the 1\e density is reduced by about 20\%; with larger $R_L$ it continues to fall, approaching the value in the cleanest regions of the cleanest quadrants. The value of $R_L=30$ was adopted from a previous unpublished analysis of SENSEI data at MINOS, and validated against its performance in the SNOLAB commissioning dataset.

We performed several checks to assess the impact of the low-energy cluster mask on our analysis.
First, we implemented a Monte Carlo simulation to produce images where all pixel values were sampled from the same Poisson distribution, such that the true charge density was uniform.  We did this both for charge densities similar to those observed in our data and for much larger charge densities, where the average distance between clusters is $\ll R_L$.
We found that applying the low-energy cluster mask to these images did not change the measured density of 1\e or multi-\e events.
In another study, we removed the low-energy cluster mask after opening the hidden data and inspected the 3\e events passing cuts. Removing the low-energy cluster mask initially resulted in more observed 3\e events, but after further inspecting the images and removing 3\e events in regions affected by understood but inadequately masked detector backgrounds such as those shown in Fig.~\ref{fig:bleed_edge}, we observed no additional 3\e events beyond those reported in Table~\ref{tab:summary}. This leads us to conclude that this mask does not suppress the density of DM-like events.
We expect that in future work with a larger dataset, improvements to masking due to improved understanding of backgrounds will reduce or remove the need for this mask.

\section{Cut Flow}\label{app:cut-flow}
Table~\ref{tab:eff} lists the impact on exposure and event counts as successive masks are applied. 

\begin{table}[htb]
\begin{center}
    \begin{tabularx}{\textwidth} 
    {|X||c|c|r|r|r|r|r|}
\hline
 & Expo. & & \multicolumn{3}{c|}{Events, 2\e} & \multicolumn{2}{c|}{Events, 3\e} \\ \hline
Mask & [g-day] & Eff. &  horz & vert & diag & 2p & 3p \\
\hline
\hline
- & 534.89 & - & 60121 & 25993 & 31572 & 10759 & 41240 \\
Noise & 436.53 & 0.817 & 9109 & 10623 & 9584 & 5636 & 9176 \\
Crosstalk & 435.68 & 0.998 & 9031 & 10226 & 9525 & 5248 & 8996 \\
Edge & 379.41 & 0.871 & 7250 & 8486 & 7314 & 4392 & 7432 \\
Bad col. & 217.88 & 0.574 & 5357 & 4008 & 5512 & 2808 & 5696 \\
Bad pix. & 217.86 & 1.000 & 5347 & 4000 & 5503 & 2804 & 5691 \\
Full-well & 213.08 & 0.978 & 981 & 435 & 759 & 181 & 511 \\
Serial reg. & 213.05 & 1.000 & 789 & 335 & 557 & 139 & 333 \\
Bleed & 208.96 & 0.982 & 740 & 275 & 446 & 136 & 312 \\
Halo & 194.43 & 0.931 & 687 & 239 & 397 & 128 & 286 \\
Low-E & 100.72 & 0.522 & 188 & 13 & 32 & 3 & 7 \\
Shape & 100.72 & 1.000 & 10 & 13 & 32 & 3 & 1 \\
\hline
\end{tabularx}
\caption{Pixel exposures (``Expo.''), cut efficiencies (``Eff.''), and number of 2\e and 3\e events passing successive masking cuts, for the combined data.
The pixel exposure is simply the exposure of unmasked pixels, and does not account for the reconstruction efficiency resulting from diffusion, misidentification, or geometric effects of particular cluster sizes or shapes.
Note that the Cluster Shape mask reduces the reconstruction efficiency for the 2\e horizontal and 3\e 3-pixel bins in particular, which is not reflected here.
} 
\label{tab:eff}
\end{center}
\end{table}

The Bad Column and Low-Energy Cluster masks account for the majority of the cut efficiency loss.
The columns in each quadrant closest to the readout amplifier have higher 1\e density due to amplifier light, and the Bad Column mask thus removes a large fraction (from 21\% to 58\%) of each quadrant.
As discussed in Section~\ref{app:Low-E-Mask}, the Low-Energy Cluster mask removes a fraction of each image, which depends on the local density of 1\e events.

\section{Calculating Limits (Description of Likelihood Fit)}\label{app:likelihood}

For both the hidden data and the commissioning+hidden data, we combine the results from the various charge and shape bins into a combined-limit constraint by utilizing the maximum log-likelihood function, as described below. 

Let $n_i$ be the number of events observed in bin $i$, $s_i$ the number of events expected from a hypothesized signal, $b_i$ the number of events expected from unknown background, and $k_i$ the number of events expected from known backgrounds, where each of these values reflects the exposure and efficiency of the data.
For this analysis, $k_i$ consists of the expected background from pileup of single-electron events, and $s_i$ depends on the DM model (including its interaction, mass, and the type of mediator) and the effective exposure given in Table \ref{tab:summary}, which includes the reconstruction efficiency.

The number of expected events in a bin is $\mu s_i+b_i+k_i$, where $\mu$ is a weighting factor for the signal hypothesis that is proportional to the cross section ($\mu=0$ corresponds to the background-only hypothesis, while $\mu=1$ is the test hypothesis that corresponds to a reference cross section of $\overline\sigma_e=10^{-36}$~cm$^2$). 

The likelihood function for each bin is the Poisson probability,
\begin{equation}
    L_i(\mu;b_i)=\frac{(\mu s_i+b_i+k_i)^{n_i}}{n_i!}\exp[-(\mu s_i+b_i+k_i)]\, .
\label{eq:likelihoodfunction}
\end{equation}
The combined likelihood function is the product over all bins, 
\begin{equation}
    L(\mu;\mathbf{b})=\prod_i L_i(\mu;b_i)\, ,
\end{equation}
where $\mathbf{b}=(b_1,b_2,b_3,\dots)$.
To test the hypothesized value of $\mu$, we define the profile likelihood ratio
\begin{equation}
\label{eq:lambda}
    \lambda(\mu)=\frac{L(\mu;  \mathbf{\Hat{\Hat{b}}})}{L(\hat{\mu};\mathbf{\hat{b})}}\, .
\end{equation}
Since we are not working with a background model, we set the (unknown) value of $\mathbf{b}$ as a nuisance parameter and marginalize over $\mathbf{b}$. 
Here, $\mathbf{\Hat{\Hat{b}}}$ denotes the value of $\mathbf{b}$ that
maximizes $L$ for the specified value of $\mu$, while $\hat{\mu}$ and
$\mathbf{\hat b}$ are the values that maximize $L$ (but constraining 
$\hat{\mu}$ to be non-negative, since the 
presence of the signal can only {\it increase} the event rate). 
According to~\cite{Cowan:2010js}, setting $\mathbf{b}$ as a nuisance parameter that depends on $\mu$ reflects the loss of information about $\mu$ due to systematic uncertainties.
Therefore, we define as test-statistic 
\begin{equation}
    \tilde t_\mu=-2\ln\lambda(\mu)= -2\ln\frac{L(\mu;  \mathbf{\Hat{\Hat{b}}}(\mu))}{L(\hat{\mu};\mathbf{\hat{b})}} \,.
\end{equation}

The corresponding $p$-value for a hypothesis $\mu$ is 
\begin{equation}
    p(\mu)=\int_{\tilde t_{\mu,{\rm obs}}}^\infty f(\tilde t_\mu|\mu) d\tilde t_\mu\, ,
\end{equation}
where $f(\tilde t_\mu|\mu)$ is the PDF of $\tilde t_\mu$ under the assumption of signal strength $\mu$.

We compute $f(\tilde t_\mu|\mu)$ using a toy-Monte Carlo (MC). 
In our particular case, the global and restricted maximums ({\it e.g.}, the denominator and numerator of Eq.~\ref{eq:lambda}) can be computed analytically. Specifically, the global likelihood (denominator) is always maximized for $\mathbf{s}=0$, $\mathbf{b}=\mathbf{n-k}$. The MC is used to generate the distribution of the test-statistic for a fixed set of parameters by Poisson-fluctuating the predicted number of events for that set of parameters. 
For each $\mu$, we generate a synthetic set of fake experiments from the parameters $\mu$ and $\mathbf{\Hat{\Hat{b}}}(\mu)$ by Poisson-fluctuating the predicted number of events, $\mu \mathbf{s}+ \mathbf{b} + \mathbf{k}$, for that set of parameters. We then compute $t_\mu$ for all of the synthetic experiments and generate the $t_\mu$-distribution that corresponds to this case. 
An iterative procedure is used to find the value of $\mu_{lim}$ for which $p(\mu_{lim})=0.1$. 
(We note that there is a degeneracy between $\mu s_i$ and $b_i$, and the global likelihood will remain unchanged as one increases $\mu$ (which is compensated for by decreasing $b_i$) until the number of events $\mu s_i$ in any bin $i$ exceeds $n_i-k_i$.)

For solar-reflected DM, the expected number of events, $s_i$, does not scale linearly with cross section, since the expected DM flux from the Sun itself depends on the cross section. For setting limits on solar-reflected DM, the likelihood function in Eq.~\ref{eq:likelihoodfunction} is thus modified to
\begin{equation}
    L_i(\sigma;b_i)=\frac{(s_i(\sigma)+b_i+k_i)^{n_i}}{n_i!}\exp[-(s_i(\sigma)+b_i+k_i)]\, ,
\end{equation}
where the expected number of solar-reflected DM events $s_i$ now also depends non-linearly on the hypothesized cross section $\sigma$. The limit on the DM cross section $\sigma_{\rm lim}$ is then found in a similar way to the 
methods described above.

\begin{figure}[t]
    \centering
    \includegraphics[width=0.83\textwidth]{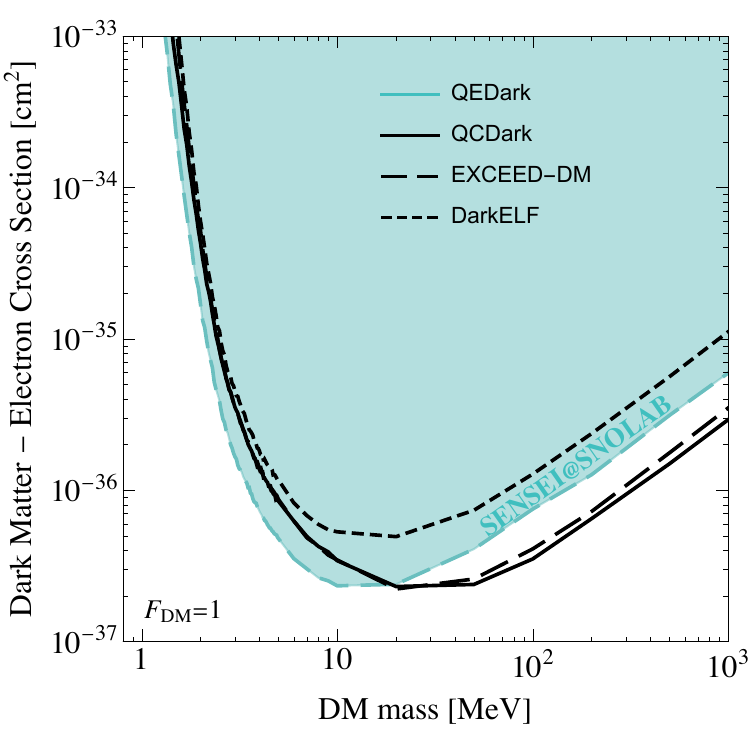}
    \includegraphics[width=0.83\textwidth]{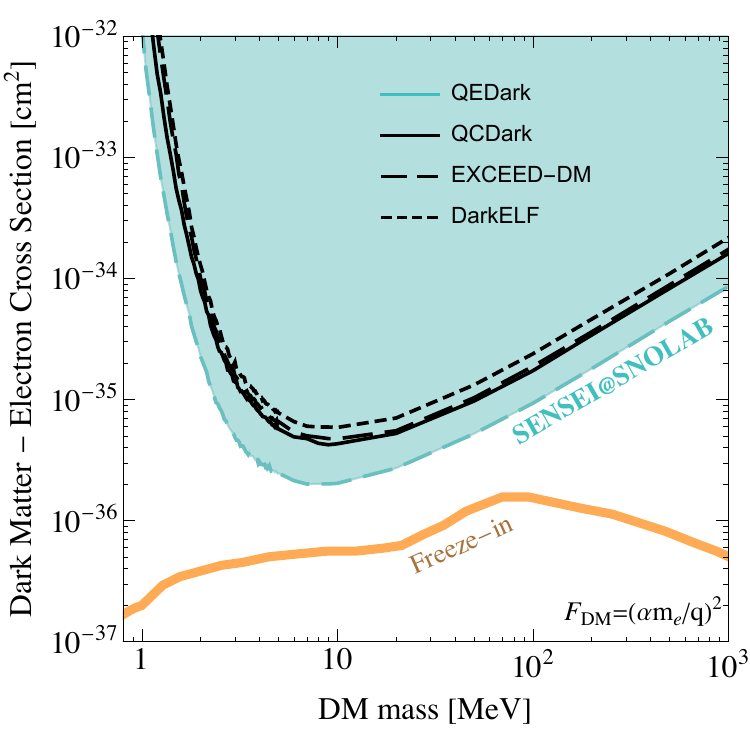}
    \caption{A comparison of the SENSEI at SNOLAB combined-data limits at 90\%~C.L.~on DM-electron scattering using different calculations for the DM-electron scattering cross section. We show the limits for a heavy mediator in the \textbf{top} panel, while the limits for a light mediator are shown in the \textbf{bottom} panel.  The cyan line uses {\tt QEDark}~\cite{Essig:2015cda} (as in the main part of the paper), the black solid line uses {\tt QCDark}~\cite{Dreyer:2023ovn,QCDark}, the long-dashed line uses {\tt EXCEED-DM}~\cite{Griffin:2021znd,Trickle:2022fwt}, and the short-dashed line uses {\tt DarkELF}~\cite{Knapen:2021bwg}.}
    \label{fig:limitcomparisons}
\end{figure}

\section{Comparison of Limits for Dark Matter-Electron Scattering}\label{app:comparison}

Fig.~\ref{fig:limitcomparisons} shows the SENSEI at SNOLAB limits at 90\%~C.L.~on DM-electron scattering using different calculations for the DM-electron scattering cross section. These different curves should \textit{not} be taken as an uncertainty on the actual limit, seeing as the various calculations account for different effects. {\tt QEDark} and {\tt DarkELF} neglect all-electron effects and therefore produce rates that are lower than those calculated with {\tt QCDark} and {\tt EXCEED-DM} for events containing more than two electrons for a heavy mediator. Hence, the {\tt QCDark} and {\tt EXCEED-DM} limits are stronger for a heavy mediator at larger DM masses. The original version of {\tt QEDark} also neglects screening effects, which are included in the other codes, and hence over-predicts the rates at low energies and light mediators. While limits produced using {\tt QEDark} are shown in the main paper to conform with calculations from other experiments, the most accurate limits are those shown with {\tt QCDark} and {\tt EXCEED-DM}. 

\begin{figure}[t]
    \centering
    \includegraphics[width=0.83\textwidth]{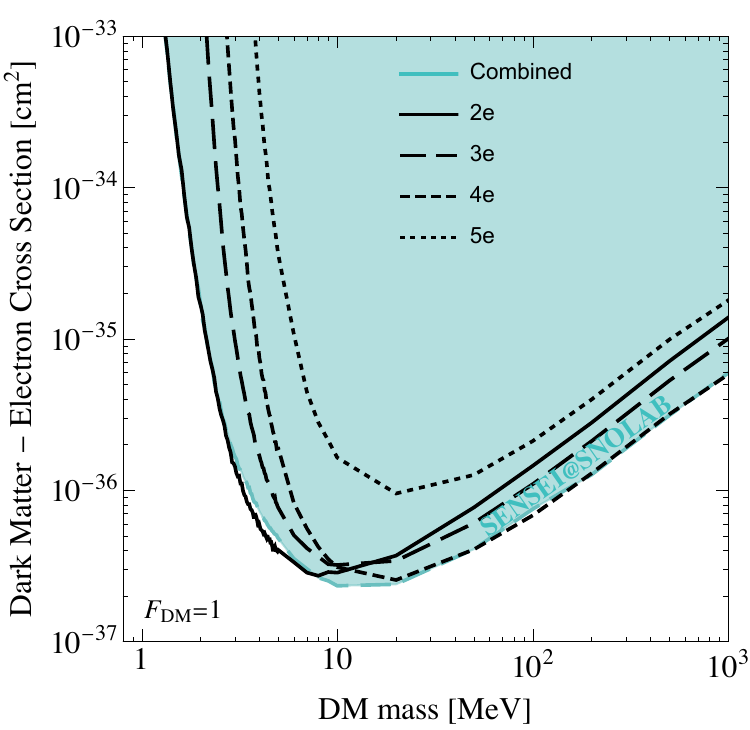}
    \includegraphics[width=0.83\textwidth]{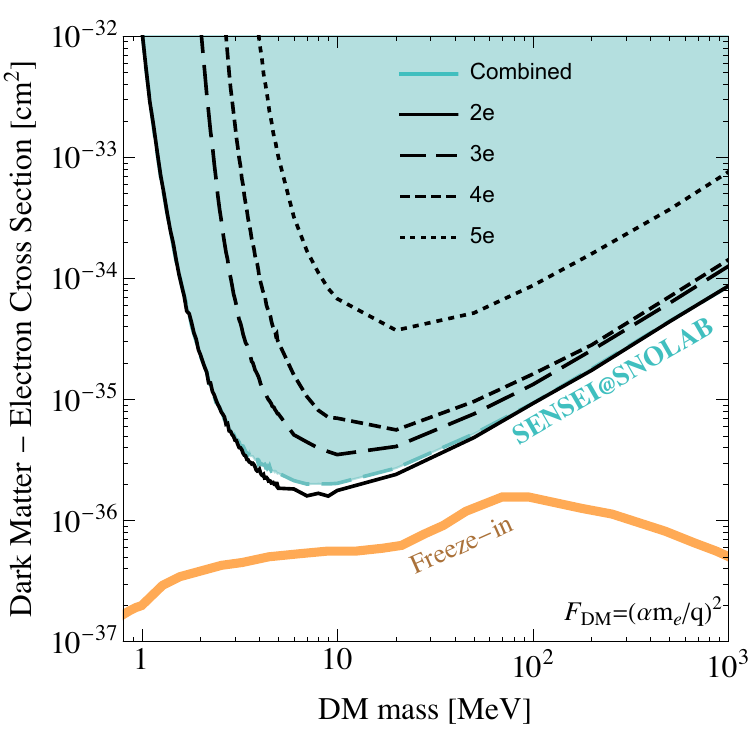}
    \caption{A comparison of the SENSEI at SNOLAB combined-data limits at 90\%~C.L.~on DM-electron scattering using the likelihood fit described in Section~\ref{app:likelihood} versus the limit from individual electron bins. We show the limits for a heavy mediator in the \textbf{top} panel, while the limits for a light mediator are shown in the \textbf{bottom} panel. }
    \label{fig:limitcomparisons-individual-vs-combined}
\end{figure}

\section{Comparison of Joint-Bin Limits with Individual-Bin Limits}\label{app:combined-vs-individual}

Fig.~\ref{fig:limitcomparisons-individual-vs-combined} shows the SENSEI at SNOLAB combined-data limits at 90\%~C.L.~on the DM-electron scattering cross section, $\overline\sigma_e$, using the likelihood fit described in Section~\ref{app:likelihood} and comparing it with the 90\%~C.L.~limit derived from individual electron bins: 2\e, 3\e, 4\e, and 5\e.  
We use the original version of \texttt{QEDark}~\cite{Essig:2015cda,QEdark} for these limits. We see that for the heavy-mediator case, the 2\e bin dominates at lower DM masses, while the 4\e bin dominates at higher masses.  Since no events were observed in the 4\e bin, the limit is statistics limited (assuming we continue to observe no events with more data). For the light-mediator case, the 2\e bin dominates for all masses.

\section{Migdal Limit for Dark Matter interacting with Nuclei through a Light Mediator}\label{app:Migdal-light}

\begin{figure}[t]
    \centering
    \includegraphics[width=0.83\textwidth]{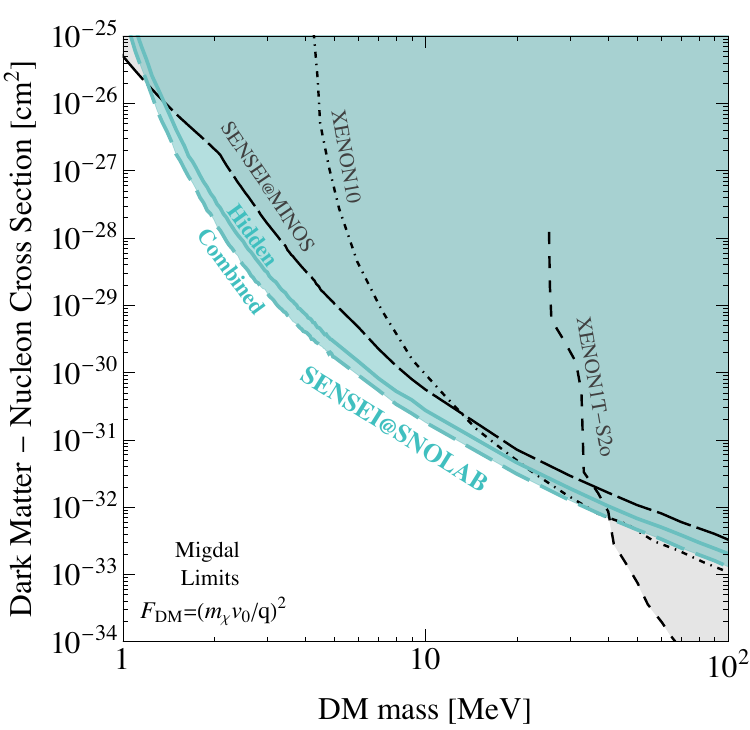}
    \caption{
    90\%~C.L.~constraints on the DM-nucleon cross section, $\overline{\sigma}_n$, for a light mediator.  
    Solid (dashed) cyan line is for the combined hidden + commissioning (hidden-only) data. SENSEI at MINOS Migdal bound has been updated from~\cite{sensei2020} using the calculations in~\cite{Berghaus:2022pbu}.  XENON10/1T Migdal bounds are from~\cite{Essig:2019xkx} using data from~\cite{Angle:2011th,Aprile:2019xxb}.}
    \label{fig:Migdal-light-mediator}
\end{figure}

Fig.~\ref{fig:Migdal-light-mediator} shows the SENSEI at SNOLAB limits at 90\%~C.L.~on the DM-nucleon scattering cross section for a light mediator, assuming the ionization signal is created via the Migdal effect.  We see that SENSEI provides the strongest bound on DM-nucleon interactions for DM masses below $\sim$40~MeV.  

\newpage

\bibliographystyle{apsrev4-1}
\bibliography{main.final.arxiv.v2.bbl}

\end{document}